\documentclass[iop,twocolappendix]{emulateapj}

\usepackage{amsmath}

\begin{document}

\title{Stacking the Invisibles: A Guided Search for Low-Luminosity Milky Way
Satellites}

\author{Branimir Sesar\altaffilmark{1,2,6}}
\author{Sophianna R.~Banholzer\altaffilmark{1}}
\author{Judith G.~Cohen\altaffilmark{1}}
\author{Nicolas F.~Martin\altaffilmark{2,4}}
\author{Carl J.~Grillmair\altaffilmark{3}}
\author{David Levitan\altaffilmark{1}}
\author{Russ R.~Laher\altaffilmark{3}}
\author{Eran O.~Ofek\altaffilmark{5}}
\author{Jason A.~Surace\altaffilmark{3}}
\author{Shrinivas R.~Kulkarni\altaffilmark{1}}
\author{Thomas A.~Prince\altaffilmark{1}}
\author{Hans-Walter Rix\altaffilmark{2}}

\altaffiltext{1}{Division of Physics, Mathematics and Astronomy, California
                 Institute of Technology, Pasadena, CA 91125,
                 USA\label{Caltech}}
\altaffiltext{2}{Max Planck Institute for Astronomy, K\"{o}nigstuhl 17, D-69117
                 Heidelberg, Germany\label{MPIA}}
\altaffiltext{3}{Spitzer Science Center, California Institute of Technology,
                 Pasadena, CA 91125, USA\label{Spitzer}}
\altaffiltext{4}{Observatoire Astronomique de Strasbourg, Universit\'e de
                 Strasbourg, F-67000 Strasbourg, France}
\altaffiltext{5}{Benoziyo Center for Astrophysics, Weizmann Institute of
                 Science, 76100 Rehovot, Israel\label{Weizmann}}
\altaffiltext{6}{Corresponding author: bsesar@mpia.de\label{email}}

\begin{abstract}
Almost every known low-luminosity Milky Way dwarf spheroidal (dSph) satellite
galaxy contains at least one RR Lyrae star. Assuming that a fraction of distant
($60 < d_{helio} < 100$ kpc) Galactic halo RR Lyrae stars are members of yet to 
be discovered low-luminosity dSph galaxies, we perform a {\em guided} search for
these low-luminosity dSph galaxies. In order to detect the presence of dSph
galaxies, we combine stars selected from more than 123 sightlines centered on 
RR Lyrae stars identified by the Palomar Transient Factory. We find that this
method is sensitive enough to detect the presence of Segue 1-like galaxies
($M_V= -1.5^{+0.6}_{-0.8}$, $r_h=30$ pc) even  if only $\sim20$ sightlines were
occupied by such dSph galaxies. Yet, when our method is applied to the SDSS DR10
imaging catalog, no signal is detected. An application of our method to
sightlines occupied by pairs of close ($<200$ pc) horizontal branch stars, also
did not yield a detection. Thus, we place upper limits on the number of
low-luminosity dSph galaxies with half-light radii from 30 pc to 120 pc, and in
the probed volume of the halo. Stronger constraints on the luminosity function
may be obtained by applying our method to sightlines centered on RR Lyrae stars
selected from the Pan-STARRS1 survey, and eventually, from LSST. In the
Appendix, we present spectroscopic observations of an RRab star in the
Bo\"{o}tes 3 dSph and a light curve of an RRab star near the Bo\"{o}tes 2 dSph.
\end{abstract}


\section{Introduction}\label{introduction}

One of the predictions of the $\Lambda$ Cold Dark Matter ($\Lambda$CDM) model is
an abundance of low-mass dark matter subhalos orbiting their host galaxies at
the present epoch \citep{kly99, mor99}. Taking into account sensitivity limits
of searches based on the Sloan Digital Sky Survey data (SDSS; \citealt{yor00}),
\citet{tol08} predict ``that there should be between $\sim300$ and $\sim600$
satellites within 400 kpc of the Sun that are brighter than the faintest known
dwarf galaxies.'' (One of the least luminous known Milky Way satellite dwarf
galaxies is Segue 1 \citep{bel07} with $M_V=-1.5^{+0.6}_{-0.8}$ \citep{mar08}.) 
While Milky Way satellites brighter than $M_V\sim-4$ have all likely been
discovered in the SDSS footprint within $\sim100$ kpc, less luminous satellites
($M_V\gtrsim-4$) may exist beyond 45 kpc from the Sun, just below the detection 
limit of current surveys (e.g., see Figure 10 of \citealt{kop08}).

To illustrate the difficulties of detecting, for example, a faint Segue 1-like
satellite at 60 kpc, consider the fact that Segue 1 has only 8 stars above the
main sequence turnoff (2 horizontal branch (HB) stars and 6 red giant branch
(RGB) stars; \citealt{sim11}). At 60 kpc, an imaging survey with a faint limit
of $r\sim22.5$ (e.g., SDSS or PS1; \citealt{met13}) could at best see these 8
stars in the satellite galaxy. Identifying these 8 stars as a statistically
significant spatial overdensity of sources in the sea of foreground stars and
background galaxies, is likely out of reach even for the most recent detection
algorithms (e.g., \citealt{wal09}\footnote{Whom we thank for inspiring the title
of this manuscript.}, \citealt{mar13}) and current datasets, unless accurate
distances or additional data are available (e.g., kinematics, chemical
abundances).

While low-luminosity (e.g., similar to Segue 1) Milky Way satellites may not be
detectable in {\em blind} searches until surveys such as the Large Synoptic
Survey Telescope (LSST; \citealt{ive08}) provide deeper multi-color imaging
covering large areas on the sky, they may be detectable in {\em guided}
searches. For example, if one had an indication of a faint satellite's location,
one could do targeted deep imaging to reach below the satellite's main sequence
turnoff and achieve a reliable detection. 

We argue that $ab$-type RR Lyrae stars (hereafter, RRab; \citealt{smi04}),
located in the outer Galactic halo (i.e., at galactocentric distances
$R_{GC}>30$ kpc), are the best practical indicators of the locations where
distant and low-luminosity Milky Way satellites may exist. As Table 4 of
\citet{boe13} and our Appendix~\ref{appendixA} show, {\em almost every
low-luminosity ($M_V\gtrsim-8$) Milky Way dwarf satellite galaxy has at least
one RRab star}. Even the least luminous of Milky Way dwarf satellites, Segue 1,
has one RR Lyrae star \citep{sim11}. This finding shows that RR Lyrae stars are
plausible tracers of even the least luminous and most metal-poor Milky Way dwarf
satellites.

RR Lyrae stars have several properties that make them useful as tracers of halo
structures. First, they are bright stars ($M_r=0.6$ mag at ${\rm [Fe/H]}=-1.5$
dex) that can be detected at large distances (5-120 kpc for $14 < r < 21$).
Second, distances of RRab stars measured from optical data are precise to
$\sim6\%$ \citep{ses13b} (vs., e.g., 15\% for K giants; \citealt{xue14}), and
can be improved to better than $3\%$ using infrared data \citep{kle11}. And
finally, RRab stars have distinct, saw-tooth shaped light curves which make them
easy to identify given multi-epoch observations (peak-to-peak amplitudes of
$\sim1$ mag in the $r$-band and periods of $\sim0.6$ days).

In this paper, we describe a statistical approach that amounts to a guided
search for low-luminosity Milky Way dwarf satellite galaxies, using distant
($R_{GC}>60$ kpc) RRab stars as indicators of their position. The sample of RRab
stars is described in Section~\ref{PTF_RR}). Instead of attempting to detect an
overdensity of sources (i.e., a low-luminosity dSph) at a particular position
and distance indicated by an RRab star, we search for evidence of faint dSphs in
the ensemble color-magnitude diagrams, stacked at the positions and distances
of RR Lyrae stars. Our detection method is described in Sections~\ref{method}
and~\ref{detection_significance}, and its sensitivity is measured in
Section~\ref{sensitivity}. The application of our method to SDSS DR10 imaging
data \citep{ahn14} is described in Section~\ref{results}, and the results are
discussed in Section~\ref{discussion}.

\section{RR Lyrae Stars}\label{PTF_RR}

RRab stars used in this work were selected by an automated classification
algorithm that uses imaging data provided by the Palomar Transient Factory
survey (PTF). Below we briefly describe the PTF survey and the RR Lyrae
selection procedure.

The PTF\footnote{\url{http://www.ptf.caltech.edu}} \citep{law09,rau09} is a
synoptic survey designed to explore the transient sky. The project utilizes the
48-inch Samuel Oschin Schmidt Telescope on Mount Palomar. Each PTF image covers 
7.26 deg$^2$ with a pixel scale of $1.01\arcsec$. The typical PTF cadence
consists of two 60-sec exposures separated by $\sim1$ hour and repeated every
one to five days. By June 2013, PTF observed $\sim11,000$ deg$^2$ of sky at
least 25 times in the Mould-$R$ filter\footnote{The Mould-$R$ filter is similar 
in shape to the SDSS $r$-band filter, but shifted {27 \AA} redward.} (hereafter,
the $R$-band filter), and about 2200 deg$^2$ in the SDSS $g^\prime$ filter. PTF
photometry is calibrated to an accuracy of about 0.02 mag \citep{ofe12a,ofe12b} 
and light curves have relative precision of better than 10 mmag at the bright
end, and about 0.2 mag at the survey limiting magnitude of $R=20.6$ mag. The
relative photometry algorithm is described in
\citet[see their Appendix A]{ofe11}.

Briefly, to select RR Lyrae stars from PTF we first searched for variable PTF
sources that have SDSS colors consistent with colors of RR Lyrae stars
(Equations 6 to 9 of \citealt{ses10}). A period-finding algorithm was then
applied to light curves of color-selected objects, and objects with periods in
the range 0.2-0.9 days were kept. Light curves were phased (period-folded) and
SDSS $r$-band RR Lyrae light curve templates (constructed by \citealt{ses10})
were fitted to the phased data. A Random Forest classifier, trained on the
sample of RR Lyrae stars identified in SDSS Stripe 82 \citep{ses10}, was then
used to classify candidate RR Lyrae stars observed by PTF (Banholzer et
al.~2014, in prep). Initial tests indicate that the samples of RR Lyrae stars
selected by our classification algorithm are highly pure ($\gtrsim95\%$) and at
least 95\% complete within 80 kpc.

We note that the completeness tests were done assuming $R=20.6$ mag as the
$5\sigma$ detection limit for PTF \citep{law09}, which assumes $2\arcsec$ seeing
and no clouds. Since not all PTF observations were done in such conditions, the
true completeness will vary as a function of the sky position. For example, PTF
fields observed during summer months will be deeper and have better photometry
than fields observed during winter months, when the atmospheric conditions are
less favorable. A completeness map for RRab stars identified by PTF that takes
into account various selection effects (survey cadence, depth of specific
fields), will be released in the upcoming paper (Banholzer et al.~2014, in
prep).

For this work, we limit our sample of RRab stars to those with heliocentric
distances greater than 60 kpc, for three reasons. First, halo structures (e.g.,
dSphs) located at greater heliocentric distances are more likely to remain
spatially coherent for a longer time due to longer orbital periods. Second, the
number density profile of RR Lyrae stars declines more steeply beyond $\sim30$
kpc from the Galactic center \citep{wat09, ses10}. The consequence of this
steepening is a reduction in the number of RR Lyrae stars beyond 30 kpc that are
likely associated with the smooth stellar spheroid (e.g., see Figure 11 of
\citealt{ses10} for an illustration). By considering only distant RR Lyrae
stars, we minimize the number of RR Lyrae stars that are likely associated with
the smooth halo (and not associated with potential dSphs) and therefore minimize
the background in our search. And third, according to Tables 2 and 3 of
\citet{kop08}, blind searches based on SDSS data have likely found all Segue
1-like dSphs within $\sim45$ kpc from the Sun. Thus, further searches in the
SDSS footprint and within 45 kpc from the Sun are not likely to uncover new
dSphs.

\begin{deluxetable}{rrr}
\tabletypesize{\scriptsize}
\setlength{\tabcolsep}{0.02in}
\tablecolumns{3}
\tablewidth{0pc}
\tablecaption{Positions and distances of RR Lyrae stars\label{table1}}
\tablehead{
\colhead{R.A.} & \colhead{Dec} & \colhead{Helio.~distance$^a$} \\
\colhead{(deg)} & \colhead{(deg)} & \colhead{(kpc)}
}
\startdata
1.242728 &  3.098036 & 88.8 \\
1.682986 & 21.730624 & 72.9 \\
3.939427 & -5.579277 & 78.4
\enddata
\tablenotetext{a}{Calculated assuming $M_r = 0.6$ as the absolute magnitude of
RR Lyrae stars in the PTF $R$-band. The fractional uncertainty in distance is
6\%.}
\tablecomments{Table~\ref{table1} is published in its entirety in the electronic
edition of the Journal. A portion is shown here for guidance regarding its form
and content.}
\end{deluxetable}

\begin{figure*}
\plottwo{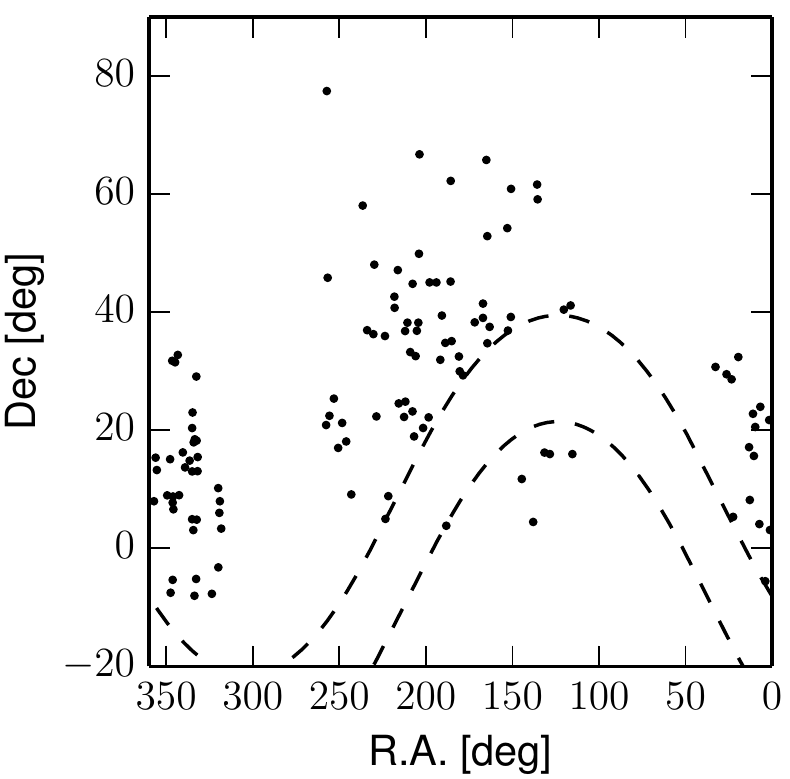}{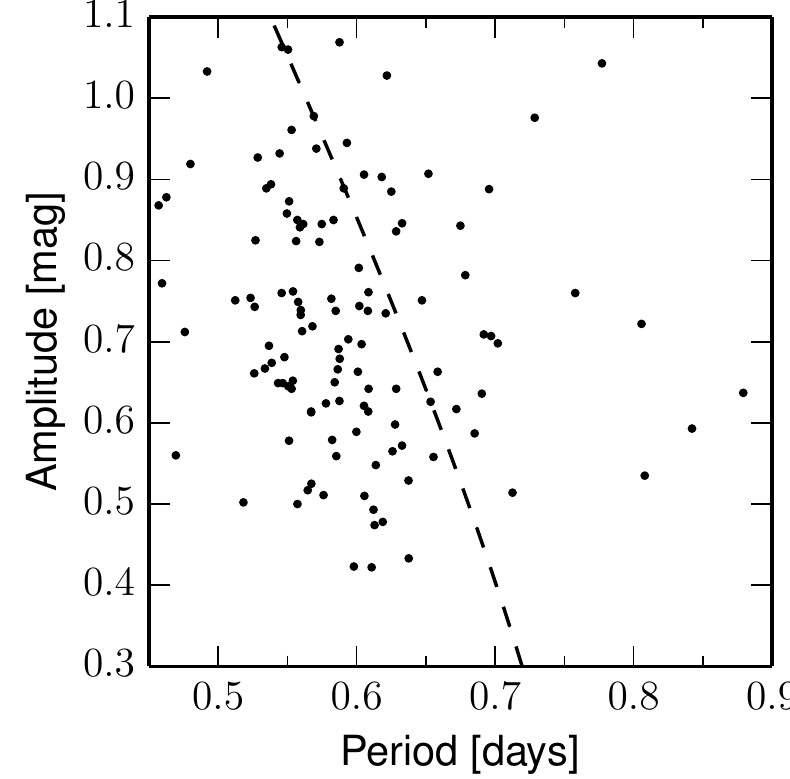}
\caption{
{\em Left:} The angular distribution of RRab stars used in this work. The dashed
lines show $\pm9\arcdeg$ off the Sagittarius stream orbital plane. There are
about 60 RRab stars within this region that were not included in our sample.
{\em Right:} The distribution of selected RRab stars in the period vs.~amplitude
diagram. The dashed line (defined by \citealt{ses13a}, see their Section 4)
separates Oosterhoff type I (short-period) from type II (long-period) RRab
stars.
\label{fig1}}
\end{figure*}

In total, our sample consists of 123 RRab stars that cover 9000 deg$^{-2}$ of
sky and are located within 60 to 100 kpc from the Sun. Their distribution in
equatorial coordinates is shown in Figure~\ref{fig1} ({\em left}) and their
positions are listed in Table~\ref{table1}. We removed RR Lyrae stars associated
with known dSphs or globular clusters. The RRab stars within $9\arcdeg$ off the
orbital plane of the Sagittarius stream are more likely to be associated with
the stream than with a low-luminosity dSph, and were thus excluded from our
sample. The period vs.~amplitude diagram (right panel of Figure~\ref{fig1})
shows the distribution of RRab stars according to the Oosterhoff classification
(\citealt{oos39}, or see Section 5.1 of \citealt{zin14} for a brief review). We
find the ratio of Oosterhoff type I and II RRab stars to be $4:1$. The same
ratio was also found by previous studies that used samples of closer RR Lyrae
stars \citep{mic08,dra13,ses13a,zin14}.

\section{Detection Method}

\subsection{Basis of the Method}\label{method}

\begin{figure}
\epsscale{1.0}
\plotone{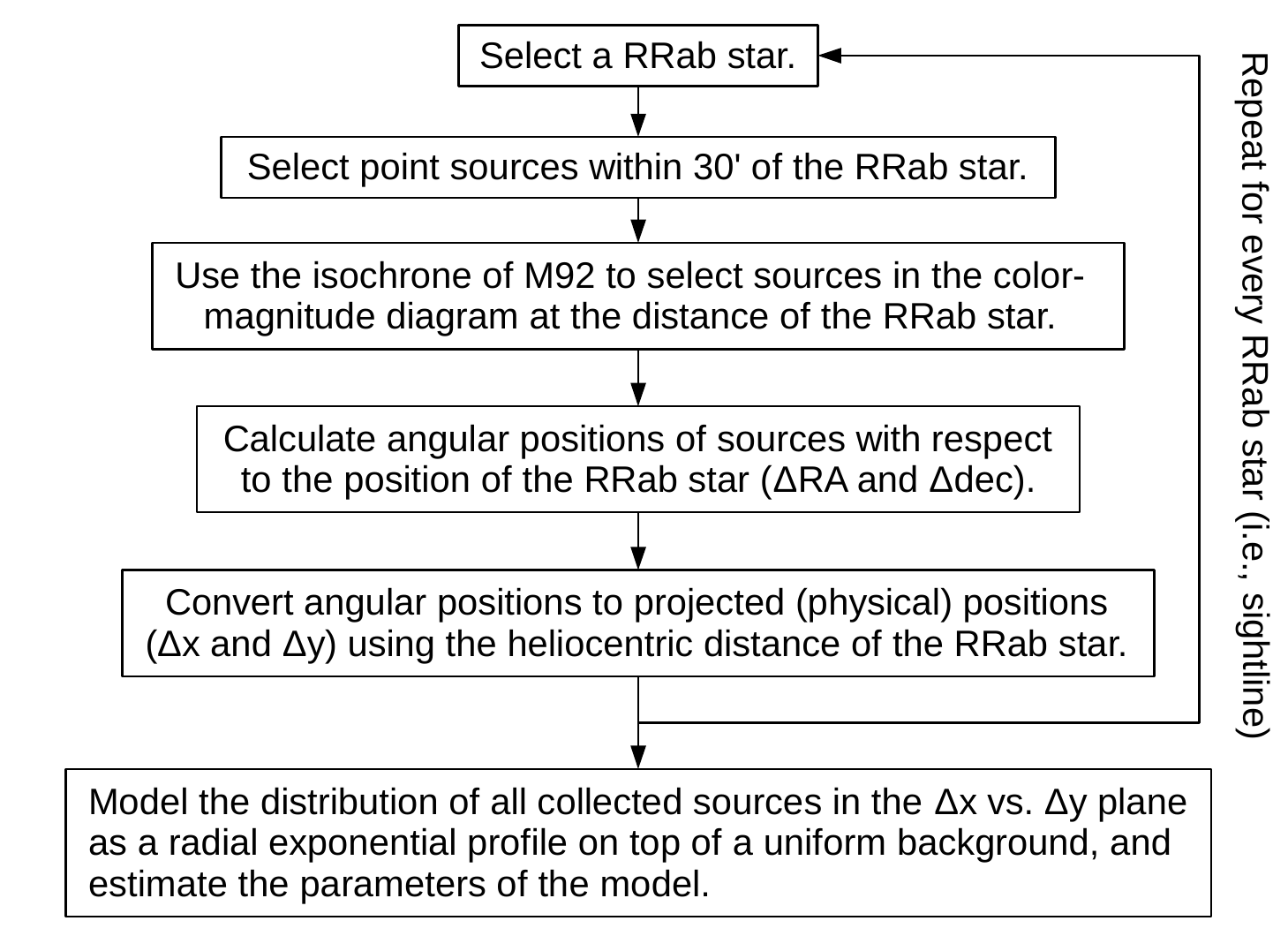}
\caption{A flowchart of our detection method.
\label{flowchart}} 
\end{figure}

While only a few HB and RGB stars may be observable in a distant low-luminosity 
dSph, these stars should be located in the vicinity of a distant RRab
star\footnote{Of course, this idea works only if RRab stars trace positions of
low-luminosity dSph in the first place. The fact that almost every known
low-luminosity dSph has an RR Lyrae star (see Section~\ref{introduction}),
supports this assumption.}. By converting the positions of stars near the RRab 
star to a coordinate system where the RRab star is in the center, and by
stacking many stars from different sightlines (all centered on different RRab
stars), one may hope to find a statistically significant overdensity of sources
near the origin of this coordinate system, indicating a statistical detection of
low-luminosity dSphs. Naturally, this signal may be detectable only if a
sufficient fraction of sightlines contains a dSph (see Section~\ref{sensitivity}
for a discussion). A flowchart of the detection method is shown in
Figure~\ref{flowchart} and described below.

First, we select a distant RRab star from our RRab sample, with a heliocentric
distance between 60 and 100 kpc. We then select all SDSS point sources brighter
than $r=21.5$ mag\footnote{In this work, we use SDSS point-spread function (PSF)
magnitudes.} and within $30\arcmin$ of the position of the RRab star. To avoid
creating an overdensity consisting only of RRab stars, we ignore sources within
$1\arcsec$ of the position of the RRab star. The SDSS morphological star-galaxy
classification is reliable for sources brighter than $r=21.5$ \citep{lup02}, and
thus we expect little or no contamination from unresolved galaxies.

To select candidate RGB stars at the distance indicated by the RRab star, we
compare positions of SDSS sources in the $g-r$ vs.~$r$ color-magnitude
diagram\footnote{From here on, all SDSS colors and magnitudes are dereddened
using dust maps of \citet{SFD98}} with a theoretical
BaSTI\footnote{\url{http://basti.oa-teramo.inaf.it/}} isochrone \citep{pie06},
shifted using the distance modulus of the RRab star (i.e., we do color-magnitude
diagram filtering; \citealt{gri95}). More specifically, we use the isochrone of
an old (12.6 Gyr), metal-poor (${\rm [Fe/H] = -2.1}$ dex), $\alpha$-enhanced
population, with an assumed mass loss parameter $\eta=0.4$.

In practice, we require that a source's $g-r$ color is within $2\sigma_{g-r}$ or
0.2 mag from the isochrone, where $\sigma_{g-r}$ is the uncertainty in $g-r$
color. Following \citet[see their Section 3.2]{wal09}, we remove sources with
$g-r>1.0$, as including redder objects adds more noise from Milky Way dwarf
stars than signal from more distant RGB stars. Similarly to candidate RGB stars,
candidate blue horizontal branch (BHB) stars are selected by comparing
dereddened $g$-band magnitudes of SDSS sources against a fiducial BHB line
defined by \citet[Equation 5]{fs13}.

The angular positions of selected sources relative to the position of the RRab
star, $\Delta RA = (RA - RA_{RRab})\cos((Dec+Dec_{RRab})/2)$ and
$\Delta Dec = Dec - Dec_{RRab}$, are converted to projected (physical) positions
$\Delta x = \Delta RA \cdot d$ and $\Delta y = \Delta Dec \cdot d$, where $d$ is
the heliocentric distance of the RRab star. Only sources within the projected
distance of $\sqrt{\Delta x^2 + \Delta y^2} < 500$ pc from the RRab star are
kept.

The above selection procedure is repeated for 123 sightlines centered on 123
RRab stars. Sources from each sightline that pass the selection are collected
and their projected positions are stored.

\subsection{Maximum Likelihood Estimation of the Significance of Detection}\label{detection_significance}

To test for the presence of an overdensity of sources near the origin of the
$\Delta x$ vs.~$\Delta y$ coordinate system, we use the maximum likelihood
approach of \citet[see their Section 2.1]{mar08}.

We model the spatial distribution of sources in the $\Delta x$ vs.~$\Delta y$
plane with an axially symmetric exponential radial density profile centered on
the origin and superposed on a uniform field contamination. Given this spatial
model, the probability of finding a data point $i$ at distance
$r_i = \sqrt{\Delta x_i^2 + \Delta y_i^2}$ from the origin is
\begin{equation}
    P_i(r_i | N^*, r_e) = \frac{2\pi r_i}{N_{tot}} \left( \frac{N^*}{2\pi r_e^2}\exp\left(-\frac{r_i}{r_e}\right) + \Sigma_b \right)\label{spatial_model},
\end{equation}
where $N^*$ is the number of stars (in the stack) that belong to low-luminosity
satellites, $r_e$ is the exponential scale radius of the profile (half-light
radius is $r_h=1.68r_e$), and $\Sigma_b$ is the surface density of foreground
stars (in units of stars pc$^{-2}$)
\begin{equation}
    \Sigma_b = \frac{N_{tot} - N^*}{r_{max}^2\pi}\label{sigb}.
\end{equation}
In Equations~\ref{spatial_model} and~\ref{sigb}, $N_{tot}$ is the total number
of sources in the stack and $r_{max}= 500$ pc.

Given a spatial distribution of a set ($\mathcal{D}_n$) of $N_{tot}$ points, the
likelihood of the entire data set is
\begin{equation}
    \mathcal{L}\left(\mathcal{D}_n | N^*, r_e\right) = \prod_{i=1}^{N_{tot}} P_i(r_i | N^*, r_e).\label{likelihood_data}
\end{equation}
The probability of a model given the data, $P(N^*, r_e)$, is then
\begin{equation}
    P(N^*, r_e | \mathcal{D}_n) \propto \mathcal{L}\left(\mathcal{D}_n | N^*, r_e\right) P(N^*)P(r_e),\label{probability_model}
\end{equation}
where $P(N^*)=1/2000$ for $0 < N^*/{\rm stars} < 2000$ and
$P(r_e)=\mathcal{N}(\log(r_h=1.68r_e) | \mu=2.26, \sigma=0.8)$ is a normal
distribution in $\log(r_h=1.68r_e\, {\rm pc^{-1}})$, centered on $\mu=2.26$ and 
with a standard deviation of $\sigma=0.8$ ($P(r_e)=0$ for $r_e \leq 0$ pc).

We chose a flat prior for $N^*$ because we expect $\lesssim10$ RGB and HB stars
per sightline and have 123 sightlines in the stack (a sightline with a dSph that
has much more than 10 stars would likely be luminous enough to be detected by
now). Thus, the expected number of dSph stars in the stack ($N^*$) may range
from 0 to $\lesssim2000$ ($\sim10\times123\lesssim2000$).

The prior probability $P(r_e)$ has been chosen based on the size-luminosity
relation for Milky Way dSph satellites \citep{bra11}. The center $\mu$ of the
normal distribution $\mathcal{N}(\log(r_h))$ was calculated using Equation 9 of
\citet{bra11} and assuming $M_V=-2$. We have adopted a wider normal distribution
than \citet{bra11} (0.8 dex vs 0.2 dex) to account for smaller and less luminous
dSphs (i.e., similar to Segue 1) that could have been biased against in the
\citet{bra11} study (due to the surface brightness detection limit of SDSS).

The probability of the model described by Equation~\ref{probability_model} is
evaluated on a grid of ($N^*$,~$r_e$) values. The pair of ($N^*$,~$r_e$) values
that yields the highest value of $P(N^*, r_e | \mathcal{D}_n)$ represents the
model most favored by data. However, more important is to know whether this
model is significantly better than the model that contains no low-luminosity
galaxies in the stack (i.e., the model where $N^* \rightarrow 0$). As discussed
by \citet[see their Section 3.3.3]{mar13}, this information hinges on the
probability of the model marginalized over all parameters but $N^*$, or in our
case, marginalized over the grid in $r_e$,
\begin{equation}
    P_{N^*}\left(N^*|\mathcal{D}_n\right) \propto \int_{1\, pc}^{201\, pc} P\left(N^*, r_e | \mathcal{D}_n\right) dr_e.
\end{equation}

Assuming that $P_{N^*}$ follows the normal distribution, the favored model
${\rm max(P_{N^*})}$ deviates from the model with no low-luminosity dSph
galaxies by $S$ times its dispersion (i.e., a ``S-sigma detection'') for $S$
defined as \citep{mar13}
\begin{equation}
    S = \sqrt{2\ln\left(\frac{\max\left(P_{N^*}\right)}{P_{N^*}\left(N^* \rightarrow 0\right)}\right)}.
\end{equation}
In practice, we evaluate $P_{N^*}\left(N^* \rightarrow 0\right)$ at $N^* = 0.1$,
which is a small enough number for $N^*$ that it has a minimal impact on the
calculation of the significance.

\subsection{The Sensitivity of the Detection Method}\label{sensitivity}

In Sections~\ref{method} and~\ref{detection_significance}, we described a method
that aims to detect faint stellar structures (e.g., low-luminosity dSphs) by
stacking stars located near distant RRab stars. In this Section, we test the
method and quantify its sensitivity.

We test our detection method by applying it to mock catalogs of stars. These
catalogs contain mock dSphs embedded in the SDSS DR10 imaging catalog, which
serves as a realistic foreground and background for mock dSphs. The mock dSphs
are created by randomly drawing a fixed number of stars from a synthesized
old, metal-poor population. The creation of mock dSph galaxies is described in
detail in Appendix~\ref{appendixB}.

In total, 123 sightlines are generated, out of which a user-defined fraction
($f_{dSph}$) contain a mock dSph, while the remaining sightlines contain only
sources from SDSS DR10. The selection and stacking of sources is applied to all
sightlines, as described in Section~\ref{method}, with one modification. Even
though the exact distance to each mock dSph is known, this information is not
used fully when doing the color-magnitude diagram filtering. To simulate the
fact that in reality the distance modulus of RRab stars is uncertain at the 0.13
mag level ($\sim6\%$ uncertainty in distance), a new distance modulus is drawn
from a normal distribution that has a standard deviation of 0.13 mag and the
mean equal to the true distance modulus. This distance modulus is then used when
selecting sources in the $g-r$ vs.~$r$ color-magnitude diagram.

\begin{figure}
\plotone{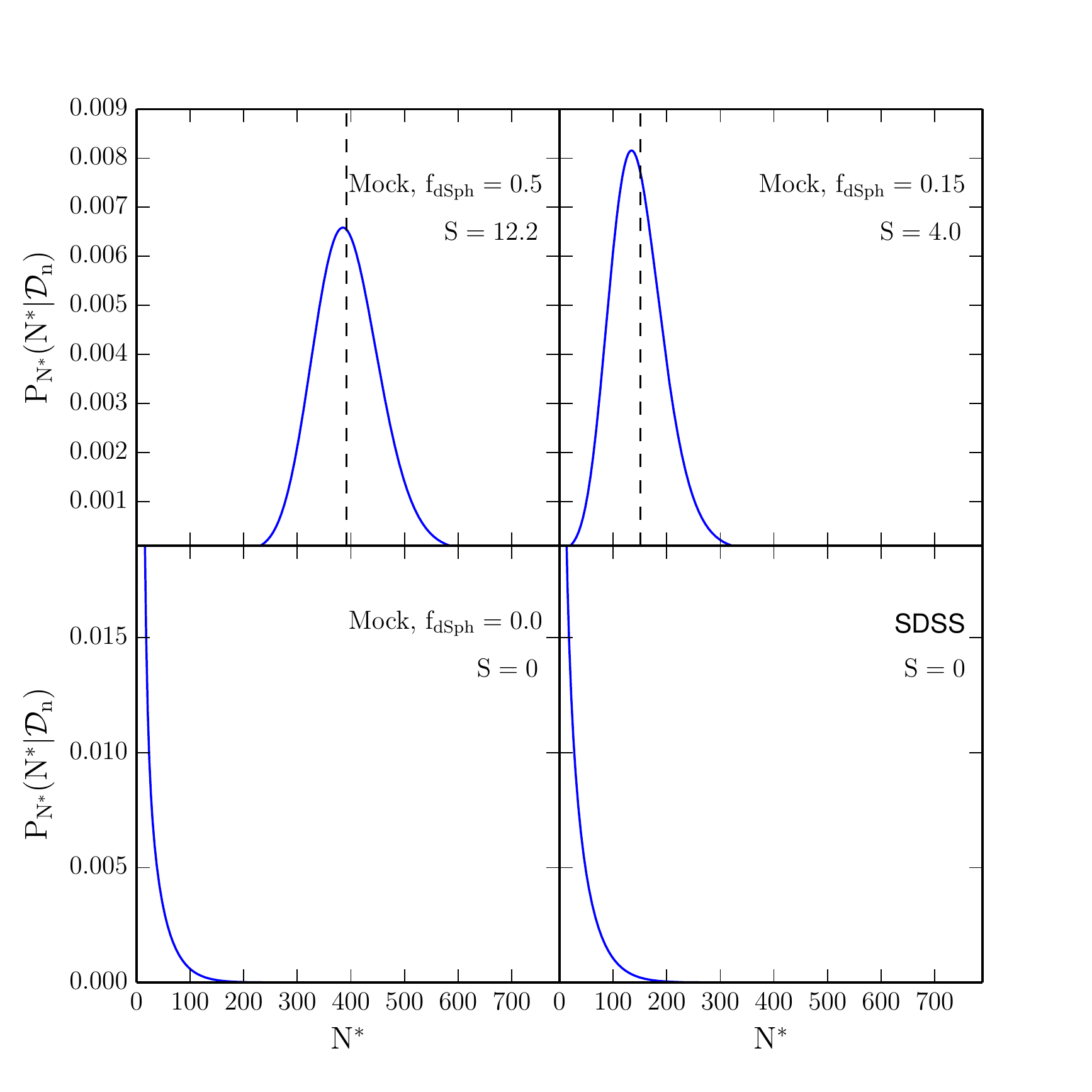}
\caption{
Marginal likelihood functions $P_{N^*}$ for three mock catalogs with different
fractions of dSphs ($f_{dSph}$), and for the SDSS DR10 catalog. The half-light
radius of dSphs used in these mock catalogs is $\sim30$ pc and their luminosity
is $M_V=-1.5^{+0.6}_{-0.8}$. The significance $S$ of detection is noted in each
panel. For comparison, the dashed line shows the true $N^*$, measured from the
mock sample with the background sources removed. Note the excellent agreement
between the true value and the value of $N^*$ where $P_{N^*}$ peaks (i.e., the
value of $N^*$ favored by data).
\label{L_Nstar}}
\end{figure}

The performance of our detection method is demonstrated by Figure~\ref{L_Nstar},
which shows the marginal likelihood functions for three mock catalogs with
different values of $f_{dSph}$. In these mock catalogs, the sightlines were
populated by dSphs in order of the heliocentric distance, from the closest to
the furthest (i.e., our ``optimistic'' scenario). Even though only a few stars
are observable in each mock dSph (the vast majority are foreground stars), and
even when only 18 out of 123 sightlines have a mock dSph ($f_{dSph} = 0.15$, top
right panel), the favored model ($N^* \sim 150$) deviates from the model with no
dSph galaxies by $4\sigma$ ($S = 4.0$), indicating a statistically significant
detection.

While the results shown in Figure~\ref{L_Nstar} are encouraging, it is important
to remember that the mock dSphs are based on Segue 1, which has a half-light
radius $r_h\sim30$ pc \citep{mar08}. Distant dSphs that have a greater
half-light radius than Segue 1 will be more difficult to detect. To test the
sensitivity of our method to dSphs with the luminosity of Segue 1
($M_V=-1.5^{+0.6}_{-0.8}$), but with a half-light radii greater than that of
Segue 1 ($r_h \sim 30$ pc; \citealt{mar08}), we increase the size of mock dSph
by some factor, create new mock catalogs, and then re-apply the detection
method.

\begin{figure}
\plottwo{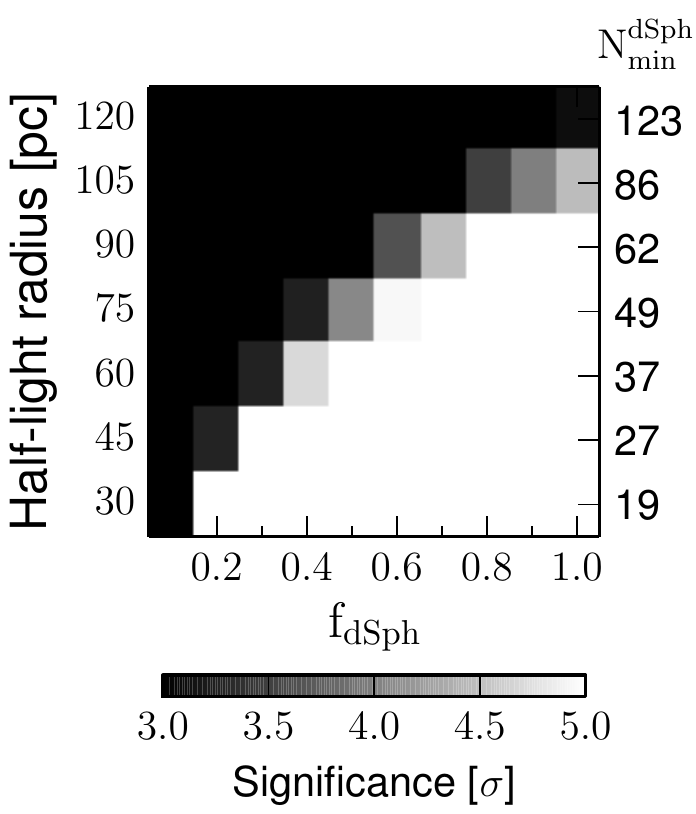}{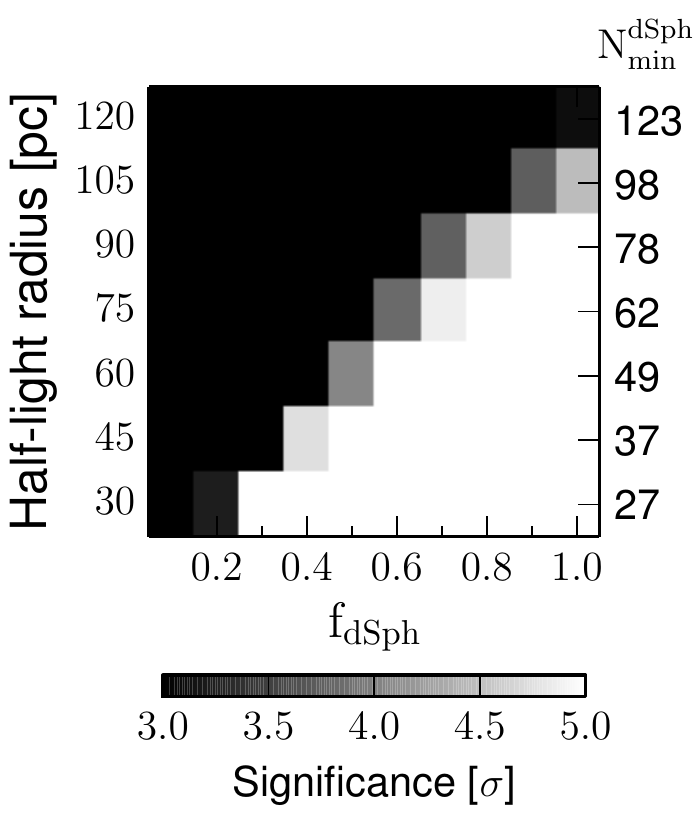}
\caption{
These panels show the regions of the parameter space where our method may or may
not detect a signal, depending on the fraction of sightlines occupied by dSphs
($f_{dSph}$) and the half-light radius of dSphs. The significance of detection
is expressed in units of standard deviation, $\sigma$, and is shown using a gray
scale which saturates at $3\sigma$ and $5\sigma$. Black pixels (significance
$<3\sigma$) show regions of the parameters space where we do not expect to be
able to detect low-luminosity dSphs ($-2.7 < M_V < -1.5$) with high statistical
significance. The left panel shows the results under the assumption that the
closest $f_{dSph}$ fraction of sightlines are populated by dSphs (the
``optimistic'' scenario), while the right panel shows the results under the
assumption that the furthest $f_{dSph}$ fraction of sightlines are populated by
dSphs (the ``pessimistic'' scenario). The column of numbers next to each panel
($N^{dSph}_{min}$) lists the minimum number of dSphs needed in the stack to
produce a $3\sigma$ detection, as a function of the half-light radius of dSphs
in the stack.
\label{significance}}
\end{figure}

Figure~\ref{significance} shows how the significance of detection changes as a
function of $f_{dSph}$ and the half-light radius of dSphs, for two different
scenarios. As expected, as the half-light radius increases, the significance of
detection decreases for a fixed $f_{dSph}$. That is, for our method to be able
to detect the presence of more extended low-luminosity dSphs, more of the
sightlines need to have a dSph. Based on Figure~\ref{significance}, we expect to
be able to detect the presence of Segue 1-like low-luminosity dSphs if more than
$\sim20$ sightlines have a dSph (the bottom row of pixels in the left panel of
Figure~\ref{significance}). The presence of low-luminosity dSph with half-light
radii of $\sim120$ pc will be detected only if every sightline has a dSph. Dwarf
spheroidal galaxies with half-light radii greater than 120 pc and with
luminosities $M_V=-1.5^{+0.6}_{-0.8}$ will likely not be detected.

We have repeated the above analysis for dSphs with the luminosity of Bo\"otes
2 ($M_V=-2.7\pm0.9$), and have found the same sensitivity limits as the one
shown in Figure~\ref{significance} for Segue 1-like ($M_V=-1.5^{+0.6}_{-0.8}$)
objects. While surprising at first, this result is not unreasonable given the
uncertainties and overlap in $M_V$ of Bo\"otes 2 and Segue 1 dSphs.

\section{Application to the SDSS DR10 Imaging Catalog}\label{results}

In this Section, we describe the application of our method to the SDSS DR10
imaging catalog. Three sets of sightlines are used to select sources from this
catalog:
\begin{itemize}
\item Sightlines centered on 123 RRab stars located beyond 60 kpc from the Sun
    and $9\arcdeg$ off the orbital plane of the Sagittarius tidal stream (i.e., our main set of sightlines).
\item Sightlines centered only on Oosterhoff type II RRab stars.
\item Sightlines centered at midpoints of close ($<200$ pc) pairs of
horizontal branch (HB) stars: RRab+RRab, RRab+BHB, or BHB+BHB star.
\end{itemize}

\subsection{Main set of sightlines}

Using the positions of 123 distant RRab stars as guiding centers, we have
selected SDSS sources located in their vicinity and have repeated the stacking
procedure described in Section~\ref{method}.

As the bottom right panel of Figure~\ref{L_Nstar} shows, the model favored by
SDSS data is the one where there is no overdensity of sources at the origin of
the $\Delta x$ vs.~$\Delta y$ coordinate system ($N^* = 0$). That is, {\em the
presence of low-luminosity dSph galaxies is not detected}. The marginal
likelihood function $P_{N^*}$ for target fields (bottom right panel) is
virtually identical to the one for background fields (i.e., fields offset
$1\arcdeg$ west from RRab stars, bottom left panel).

\subsection{Sightlines with Oosterhoff II RRab stars}

As Figure~3 of \citet{boe13} shows, metal-poor (${\rm [Fe/H] < -2}$ dex)
low-luminosity dSph galaxies mainly contain Oosterhoff type II RRab stars (i.e.,
RRab stars with periods longer than 0.65 days). As our Figure~\ref{fig1} shows,
our sample is dominated by Oosterhoff type I RRab stars.

If low-luminosity dSph galaxies predominantly contain Oo II RRab stars, it makes
sense to consider a stack consisting only of sightlines centered on Oo II RRab
stars. Using Figure~\ref{fig1}, we have selected $\sim30$ of such sightlines and
have repeated our analysis. Still, no signal was detected.

\subsection{Sightlines with pairs of HB stars}

The lack of a detection could be due to a preponderence of disrupted structures 
in the Galactic halo (e.g., streams, shells, clouds; \citealt{joh08}). The
surface brightness of structures decreases as they are disrupted, and a stack
of sightlines centered on RRab stars in disrupted structures is not likely to
yield a detection.

The fraction of sightlines centered on disrupted structures may be reduced by
considering only sightlines that contain two or more RRab stars in close
proximity. Since the number density of RRab stars in the outer halo is very low,
a detection of two or more RRab stars in close proximity ($<200$ pc) could
indicate the presence of a spatially coherent structure (e.g., a low-luminosity
dSph). Similarly, a close pair consisting of an RRab and a BHB star or a pair of
BHB stars, could serve the same purpose. Dwarf spheroidal galaxies are known to
have both types of horizontal branch stars (e.g., Segue 1). In addition, BHB
stars are expected to outnumber the RR Lyrae stars in metal-poor populations
(the ratio of BHB and RR Lyrae stars in the field is $\sim6:1$;
\citealt{psb91}).

We have searched for close pairs of RR Lyrae stars in our extended sample. The
extended sample includes RRab stars that are within $9\arcdeg$ off the orbital
plane of the Sagittarius tidal stream, and stars with heliocentric distances
greater than 45 kpc (the main RRab sample used so far starts at 60 kpc). We have
found only two pairs of RRab stars separated by less than 200 pc. To search for
close pairs of BHB and RRab stars, we have cross-matched our extended sample
with the catalog of photometrically-selected BHB stars of \citet{smi10}. We have
found 7 close pairs consisting of a BHB and an RRab star. A cross-match of the
BHB catalog with itself yielded 5 close pairs of BHB stars. We have applied our
method to the 14 sightlines described above, but once again, no signal was
detected.

We have experimented with various modifications to our detection method in order
to see if a signal appears. We have allowed fainter sources to be considered by
changing the magnitude cut from $r=21.5$ to $r=22.5$ mag. Instead of using a
single isochrone for color-magnitude diagram filtering, we adopted five BaSTI
isochrones that span a range of metallicities, from ${\rm [Fe/H] = -1.6}$ dex to
${\rm [Fe/H] = -3.6}$ dex. In the end, none of the modifications significantly
changed the bottom right panel of Figure~\ref{L_Nstar} or our main conclusion --
the presence of low-luminosity dSph galaxies is not detected.

\section{Discussion And Conclusions}\label{discussion}

Almost every known low-luminosity Milky Way dSph satellite galaxy contains at
least one RR Lyrae star (Table~4 of \citealt{boe13}). This observation, and the
fact that RR Lyrae stars can be easily identified in multi-epoch imaging,
motivated us to do a {\em guided} search for low-luminosity dSph galaxies, using
distant RRab stars as tracers of their possible locations.

We use positions and distances of RRab stars to select SDSS stars that may be
located in their vicinity (most likely, RGB stars). Sky positions of selected
stars are transformed to a physical coordinate system that is centered on RRab
stars, and sources from multiple sightlines are collected into a stack. If a
fraction of sightlines contains a low-luminosity dSph, an overdensity of sources
in the center of the stack should exist. By using mock catalogs, we have shown
that our method is able to detect the presence of dSph galaxies even if only
$\sim20$ sightlines contain a dSph as faint as Segue 1 ($M_V = -1.5$).

We have applied our method to 123 sightlines that contain RRab stars identified
by the Palomar Transient Factory (PTF) survey. These stars are spread over 9000
deg$^2$ of sky and span heliocentric distances from 60 to 100 kpc. However, we
have not detected a statistically significant signal that would indicate the
presence of low-luminosity dSph galaxies in the stack. Various modifications of
our method did not change this result. An analysis of sightlines centered on
close pairs of horizontal branch stars (separated by $<200$ pc), also did not
yield a detection.

\begin{deluxetable}{rrr}
\tabletypesize{\scriptsize}
\setlength{\tabcolsep}{0.02in}
\tablecolumns{3}
\tablewidth{0pc}
\tablecaption{Upper limits on the number of $-2.7 < M_V < -1.5$ dSphs within 9000 deg$^2$ and 60 to 100 kpc\label{table2} from the Sun}
\tablehead{
\colhead{$r_h^a$ (pc)} & \colhead{Optimistic} & \colhead{Pessimistic}
}
\startdata
30 & $<19$ & $<27$ \\
45 & $<27$ & $<37$ \\
60 & $<37$ & $<49$ \\
75 & $<49$ & $<62$ \\
90 & $<62$ & $<78$ \\
105 & $<86$ & $<98$ \\
120 & $<123$ & $<123$
\enddata
\tablenotetext{a}{Half-light radius of dSphs in the stack.}
\end{deluxetable}

\begin{figure}
\plotone{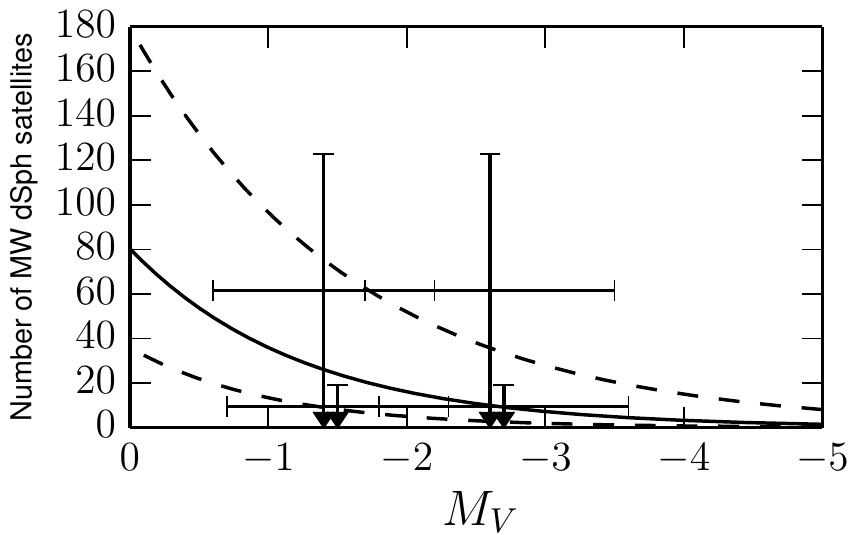}
\caption{
The solid line shows the estimated number of Milky Way dSph galaxies in the
probed volume of the halo (9000 deg$^2$ and between 60 to 100 kpc from the Sun)
based on the luminosity function of \citet{tol08}. The dashed lines show the
$1\sigma$ range of values. The symbols with errobars show the upper limits
determined in this work, for galaxies with $M_V = -1.5 \pm 0.8$ (i.e., Segue
1-like) and $M_V = -2.7 \pm 0.9$ (i.e., Bo\"otes 2-like), and with half-light
radii of $r_h=120$ pc and $r_h=30$ pc (left and right arrow, respectively, for
each $M_V$). We do not consider more luminous dSphs as such should have been
already detected within 100 kpc in SDSS data (see Figure~10 of \citealt{kop08}).
\label{fig5}
}
\end{figure}

Since no signal was detected, the sensitivity limits ($N^{dSph}_{min}$) shown in
Figure~\ref{significance} represent the upper limits on the number of
$-2.7 < M_V < -1.5$ dSphs in the 9000 deg$^2$ of sky and within 60 to 100 kpc
from the Sun (Table~\ref{table2}). In Figure~\ref{fig5}, we compare these upper
limits with the luminosity function estimate of \citet{tol08}. A comparison with
the luminosity function estimate of \citet{kop08} is not plotted, because
\citet{kop08} estimate that $<1$ low-luminosity dSph should be present in the
probed volume of the halo.

As Figure~\ref{fig5} shows, our upper limit on the number of $M_V\sim-1.5$ dSphs
with $r_h=30$ pc is slightly lower than the average estimate of \citet{tol08},
though still within their $1\sigma$ range. If dSphs with such small half-light
radii are remnants of tidally stripped galaxies, as some theoretical and
observational studies speculate \citep{pen08, br11, kir13}, then our upper limit
may constrain the efficiency and frequency of this process. The number of more
extended dSphs (i.e, $r_h=120$ pc), is progressively less constrained. This is
unfortunate, as some N-body simulations predict that the majority of
low-luminosity dSphs should have $r_h>100$ pc (e.g., \citealt{br11}).

While we were not successful in detecting low-luminosity dSphs using the current
sample of RR Lyrae stars identified by PTF, we hope to increase the volume of
the probed Galactic halo by applying our method to sightlines centered on RR
Lyrae stars identified using Pan-STARRS1 (PS1; \citealt{kai10}) multi-epoch
data. While single-epoch PS1 images are not as deep as SDSS images, we are
exploring the possibility that multi-band and multi-epoch PS1 data may be
sufficient to enable detection of RRab stars up to $\sim130$ kpc from the Sun
and over three quarters of the sky. Increasing the coverage of the PTF fields
with sufficient epochs would also help. The increase in volume should allow us
to more tightly constrain the luminosity function of Milky Way satellites and
possibly measure the degree of (an)isotropy of their spatial distribution.

The prospects of detecting low-luminosity dSphs look even more exciting, once
the multi-epoch data obtained by the Large Synoptic Survey Telescope (LSST;
\citealt{ive08}) become available. Based on realistic simulations, LSST is
expected to be able to detect $\geq90\%$ of RRab stars within $\sim360$ kpc of
the Sun across most of survey footprint, and out to $\sim760$ kpc in select deep
fields \citep{olu12}. At magnitudes corresponding to these distances
($r\sim22.8$ and $r\sim24.4$, respectively), galaxies dominate the source count
and clusters of unresolved galaxies may become a non-negligible source of
false-positive detections of dSphs. Advanced star-galaxy classification schemes
will help with this issue (e.g., \citealt{fhw12}), but actually {\em knowing}
where a dSph may be and knowing its distance will be very useful. RRab stars
observed by LSST will provide that information, hopefully for a large fraction
of yet to be discovered Milky Way satellites.

\acknowledgments

We thank \v{Z}eljko Ivezi\'c, Erik Tollerud, and Colin Slater for comments,
suggestions, and useful discussions. B.S and J.G.C thank NSF grant AST-0908139
to J.G.C for partial support, as do S.R.K (to NSF grant AST-1009987), and C.J.G 
(for a NASA grant). B.S acknowledges funding from the European Research Council 
under the European Union’s Seventh Framework Programme (FP 7) ERC Grant
Agreement n.~${\rm [321035]}$. S.R.B thanks Caltech Summer Undergraduate
Research Fellowship (SURF) for support. E.O.O. is incumbent of the Arye
Dissentshik career development chair and is grateful to support by grants from
the Willner Family Leadership Institute Ilan Gluzman (Secaucus NJ), Israeli
Ministry of Science, Israel Science Foundation, Minerva foundation, Weizmann-UK
foundation and the I-CORE Program of the Planning and Budgeting Committee and
The Israel Science Foundation. This work has made use of BaSTI web tools.

This article is based on observations obtained with the Samuel Oschin Telescope
as part of the Palomar Transient Factory project, a scientific collaboration
between the California Institute of Technology, Columbia University, Las Cumbres
Observatory, the Lawrence Berkeley National Laboratory, the National Energy
Research Scientific Computing Center, the University of Oxford, and the Weizmann
Institute of Science. It is also partially based on observations obtained as
part of the Intermediate Palomar Transient Factory project, a scientific
collaboration among the California Institute of Technology, Los Alamos National 
Laboratory, the University of Wisconsin, Millwakee, the Oskar Klein Center, the
Weizmann Institute of Science, the TANGO Program of the University System of
Taiwan, the Kavli Institute for the Physics and Mathematics of the Universe, and
the Inter-University Centre for Astronomy and Astrophysics.

Funding for SDSS-III has been provided by the Alfred P.~Sloan Foundation, the
Participating Institutions, the National Science Foundation, and the
U.S.~Department of Energy Office of Science. The SDSS-III web site is
\url{http://www.sdss3.org/}.

SDSS-III is managed by the Astrophysical Research Consortium for the
Participating Institutions of the SDSS-III Collaboration including the
University of Arizona, the Brazilian Participation Group, Brookhaven National
Laboratory, Carnegie Mellon University, University of Florida, the French
Participation Group, the German Participation Group, Harvard University, the
Instituto de Astrofisica de Canarias, the Michigan State/Notre Dame/JINA
Participation Group, Johns Hopkins University, Lawrence Berkeley National
Laboratory, Max Planck Institute for Astrophysics, Max Planck Institute for
Extraterrestrial Physics, New Mexico State University, New York University, Ohio
State University, Pennsylvania State University, University of Portsmouth,
Princeton University, the Spanish Participation Group, University of Tokyo,
University of Utah, Vanderbilt University, University of Virginia, University of
Washington, and Yale University.

\bibliographystyle{apj}
\bibliography{ms}

\appendix

\section{RRab Stars in the Bo\"{o}tes 2 and Bo\"{o}tes 3 dSph Galaxies}\label{appendixA}

Bo\"{o}tes 2 \citep{wal07} and Bo\"{o}tes 3 \citep{gri09} dSph galaxies are two
low-luminosity dSphs ($M_V=-2.7\pm0.9$ \citep{mar08} and $M_V=-5.8\pm0.5$
\citep{cor09}, respectively) that are not listed in Table 4 of \citet{boe13}
as having an RRab star. In order to verify whether these two dSphs truly lack
RRab stars, we have searched our own sample of RRab stars (selected from PTF
data) and the sample of RRab stars selected by \citet{dra13} from the Catalina 
Real-Time Sky Survey (CRTS; \citealt{dra09}).

\begin{figure}
\plotone{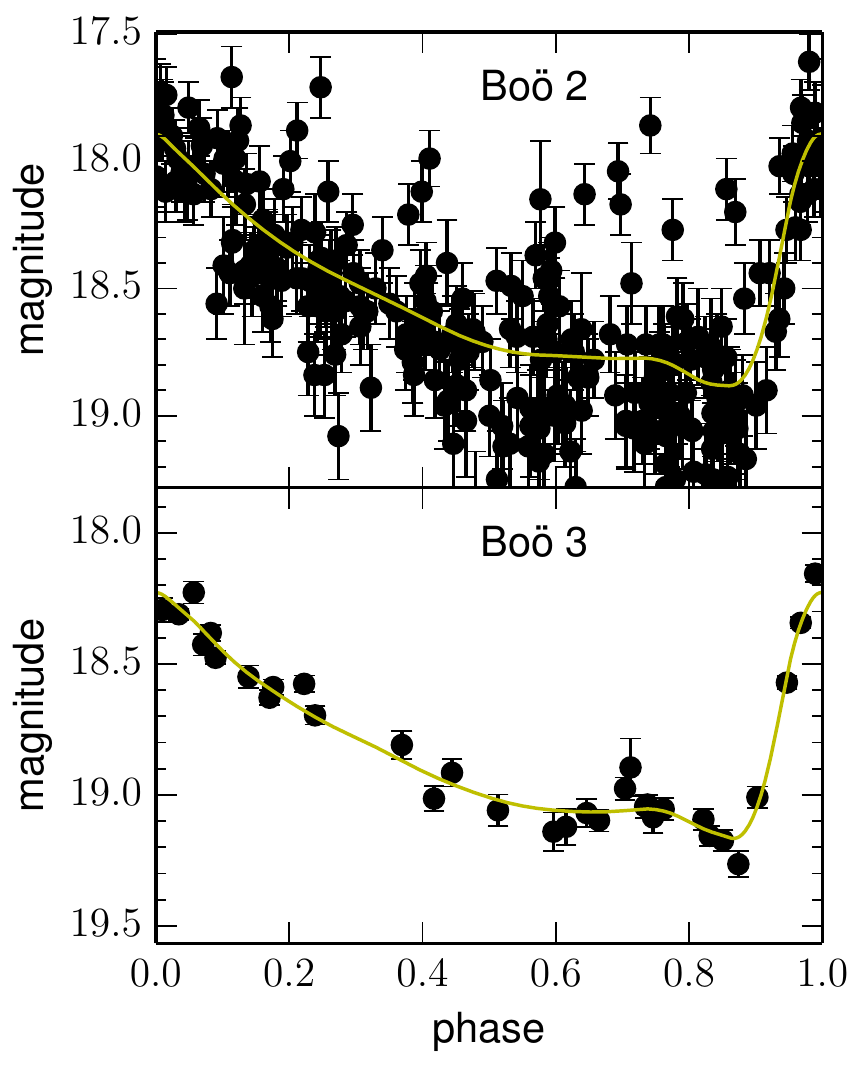}
\caption{
Phased light curves of RRab stars in the Bo\"{o}tes 2 ({\em top}) and
Bo\"{o}tes 3 dSph galaxies ({\em bottom}), observed by CRTS and PTF,
respectively. The solid lines show the best-fit $r$-band templates of
\citet{ses10}.
\label{fig6}
}
\end{figure}

We have found an RRab star near each of these dSphs and at an inferred distance 
comparable to that each of these two dSphs. Their light curves are shown in
Figure~\ref{fig6} and their light curve properties, derived from fitting
$r$-band templates of \citealt{ses10} to PTF and CRTS data, are listed in
Table~\ref{table3}.

The RRab star near Bo\"{o}tes 2 is located $39\pm2$ kpc from the Sun within
$1\arcmin.7$ off the center of Bo\"{o}tes 2. For comparison, the half-light
radius of Bo\"{o}tes 2 is $r_h = 4\arcmin.2^{1.1}_{-1.4}$
\citep{mar08}, and its heliocentric distance is $42\pm2$ kpc \citep{wal08}.
While we do not know the radial velocity of this RRab star, based on its
position and distance we conclude that it is likely associated with the
Bo\"{o}tes 2 dSph. Judging by its position in the period-amplitude diagram
(Figure~\ref{fig1}), this star is an Oosterhoff II RRab star. Using Figure~3 of
\citet{boe13}, which shows the mean RRab period vs.~mean ${\rm [Fe/H]}$ for
Milky Way dwarf galaxies with predominately old stellar populations, we estimate
that the metallicity of this RRab star may be between -2.5 and -2 dex.

The RRab star near Bo\"{o}tes 3 is located $46\pm2$ kpc from the Sun, and is
offset $\sim0\arcdeg.8$ east and south from the center of Bo\"{o}tes 3, in its
``east lobe'' (see Figure 10 of \citealt{gri09}). The heliocentric distance of
this star is equal to that of Bo\"{o}tes 3 (46 kpc; \citealt{gri09}). Based on
its position in the period-amplitude diagram (Figure~\ref{fig1}), this star is
an Oosterhoff II RRab star.

We observed this star on Aug ${\rm 2^{nd}}$ 2013, using the Double Spectrograph
(DBSP; \citealt{og82}) mounted on the Palomar 5.1-m telescope. A 600 lines
mm$^{-1}$ grating and a {5600 \AA} dichroic were used, providing a resolution of
$R=1360$ and a spectral range from 3800 {\AA} to {5700 \AA}. The velocity and
metallicity were measured following
\citet[see their Sections 2.4 to 2.7]{ses13b}.

The spectroscopic metallicity of this star, ${\rm [Fe/H]=-2.0\pm0.1}$ dex, and
its heliocentric center-of-mass velocity, $v_{helio}= 173\pm13$ km s$^{-1}$,
are fully consistent with properties of Bo\"{o}tes 3; metallicity
${\rm [Fe/H]=-2.1\pm0.2}$ dex, mean velocity $v_{helio}= 197.5\pm3.2$ km
s$^{-1}$ and velocity dispersion $\sigma_v = 14.0\pm3.2$ km s$^{-1}$
\citep{car09}. The outlying position of this RRab star (in the ``east lobe'')
and its velocity relative to the dSph (173 vs.~198 km s$^{-1}$), suggest that
the star may be part of a tidal stream extending from Bo\"{o}tes 3. If so, the
proper motion of this star, which will be measured by the Gaia mission
\citep{per01}, will be an important datum in any effort to constrain the orbit
of this dSph galaxy.

\begin{deluxetable}{lrr}
\tabletypesize{\scriptsize}
\setlength{\tabcolsep}{0.02in}
\tablecolumns{3}
\tablewidth{0pc}
\tablecaption{RR Lyrae stars in Bo\"{o}tes 2 and 3\label{table3}}
\tablehead{
\colhead{ } & \colhead{Bo\"{o} 2} & \colhead{Bo\"{o} 3}
}
\startdata
R.A.~(deg)                          & 209.52935716 & 210.14384929 \\
Dec (deg)                           &  12.85634739 & 25.93130497  \\
Survey                              &      CSS     &    PTF       \\
Period (days)                       &   0.6332816  &  0.6332751   \\
${\rm rHJD_0}$ (days)$^a$           & 55621.942645 & 55371.793948 \\
Amplitude (mag)                     &     0.99     &    0.94      \\
${\rm m_0}$ (mag)$^b$               &     17.89    &    18.23     \\
$\langle {\rm m} \rangle$ (mag)$^c$ &     18.46    &    18.80     \\
${\rm [Fe/H]}$ (dex)                & -1.79$^d$    &    -2.02     \\
$v_{helio}$ (km s$^{-1}$)$^e$       &    n.a.      &  $173\pm13$  \\
$d_{helio}$ (kpc)$^f$               &    $39\pm2$  &   $46\pm2$
\enddata
\tablenotetext{a}{Reduced Heliocentric Julian Date of maximum brightness
(HJD - 2400000).}
\tablenotetext{b}{Magnitude at maximum brightness (not corrected for
extinction).}
\tablenotetext{c}{Flux-averaged magnitude corrected for extinction.}
\tablenotetext{d}{Adopted from \citet{koc09}.}
\tablenotetext{e}{Heliocentric velocity.}
\tablenotetext{f}{Heliocentric distance calculated assuming
$M_{RR}=0.23{\rm [Fe/H]} + 0.93$ \citep{chaboyer99, cc03} as the absolute
magnitude of an RR Lyrae star.}
\end{deluxetable}

\section{Creation of Mock dSph Galaxies}\label{appendixB}

For an RRab star selected from our sample and located at $RA_0$, $Dec_0$, and
heliocentric distance $d$, a mock dSph is created as follows. First, we
randomly draw a fixed number of stars ($N_{stars}$, brighter than some absolute
SDSS $r$-band magnitude $M^{cut}_r$) from an old (12.6 Gyr), $\alpha$-enhanced,
and metal-poor (${\rm [Fe/H] = -2.3}$ dex) population generated using the
SYNTHETIC MAN population synthesis code (\citealt{cor07}; available through the
BaSTI web interface\footnote{\url{http://basti.oa-teramo.inaf.it/BASTI/WEB\_TOOLS/synth\_pop/index.html}}). The \citet{rei75} mass loss parameter for this
population is assumed to be $\eta=0.4$, and the spread in metallicity is
assumed to be ${\rm \sigma_{[Fe/H]} = 0.3}$ dex.

Following discussions by \citet[see their Section 3]{mar08} and
\citet[see their Section 3.6]{wal08}, we create mock dSphs with luminosity
$M_V$ by drawing $N_{stars}$ brighter than some absolute SDSS $r$-band magnitude
($M^{cut}_r$). For example, Segue 1 dSph ($M_V=-1.5^{+0.6}_{-0.8}$ dex) has
70 stars brighter than $M_r=4.2$ \citep{sim11}. Thus, when creating a mock dSph
with $M_V=-1.5$, we randomly draw 70 stars brighter than $M^{cut}_r=4.2$ from
the synthetic population described above. Based on the analysis of
\citet[see their Table 1]{mar08}, Bo\"{o}tes 2 dSph ($M_V=-2.7\pm0.9$) is
estimated to have 37 stars\footnote{For comparison, \citet{mar08} estimated that
Segue 1 dSph has $65\pm9$ stars brighter than $r=22$, well before \citet{sim11}
spectroscopically confirmed 70 Segue 1 stars up to the same magnitude limit.}
brighter than $M_r=3.9$ (corresponding to the magnitude limit of $r=22$ at the
heliocentric distance of 43 kpc for Bo\"{o}tes 2). Thus, when creating a mock
dSph with $M_V=-2.7$, we randomly draw 39 stars brighter than $M^{cut}_r=3.9$.

The next step is to spatially distribute drawn stars within the mock dSph. Since
the sample of Segue 1 stars observed by \citet{sim11} represents the most
complete sample of confirmed members of a dSph, we use the spatial distribution 
of stars in Segue 1 as a template for the spatial distribution of stars in our
mock dSph galaxies. To create this template spatial distribution, we first
calculate the angular positions of spectroscopically confirmed members of Segue 
1 relative to the center of Segue 1, as $\Delta RA = RA - RA_{Seg1}$ and
$\Delta Dec = Dec - Dec_{Seg1}$. These angular positions are then converted to
projected (physical) positions $\Delta x_{template} = \Delta RA \cdot d$ and
$\Delta y_{template} = \Delta Dec \cdot d$, where $d=23$ kpc is the heliocentric
distance of Segue 1.

To spatially distribute drawn stars, we randomly assign them projected
(physical) positions, $\Delta x_{template}$ and $\Delta y_{template}$, while
making sure the same position is not assigned twice. A star in the mock dSph is 
then randomly selected and designated as the RRab star of the new mock dSph. The
projected positions of other members are offset such that the mock RRab is
placed in the center of the projected coordinate system, that is,
$\Delta x = \Delta x_{template} - \Delta x_{RRab}$ and
$\Delta y = \Delta y_{template} - \Delta y_{RRab}$. This translation of
coordinates simulates the fact that for actual dSphs, the position of the RRab
star relative to the center of the dSph is not known. The $\Delta x$
vs.~$\Delta y$ coordinate system is then rotated by some random angle, and the
angular positions $\Delta RA$ and $\Delta Dec$ are calculated using the
heliocentric distance $d$ of the observed RRab star.

Finally, the mock dSph is placed $1\arcdeg$ west of the observed RRab star,
along the same galactic latitude. This placement ensures that the foreground and
background for the mock dSph are similar to the one at the position of the
observed RRab star. At the same time, by slightly offsetting the mock dSph, we
minimize the possibility of adding real dSphs into the mock catalogs, if such
exist in the SDSS DR10 catalog at the positions of observed RRab stars. 

In the final step, we modify the synthetic SDSS photometry of stars in the mock
dSph. The apparent magnitudes are calculated using the heliocentric distance $d$
of the observed RRab star and extincted using dust maps of \citet{SFD98}. The
photometric uncertainty is calculated using extincted apparent magnitudes and
models shown in Figure 1 of \citet{ses07}. The ``observed'' magnitude is then
generated by drawing a value from a normal distribution that has a standard
deviation equal to the photometric uncertainty and the mean equal to the
extincted apparent magnitude.

\end{document}